\documentstyle{article}
\newcommand\mysection{\setcounter{equation}{0}\section}
\renewcommand{\theequation}{\thesection.\arabic{equation}}
\newcounter{hran} \renewcommand{\thehran}{\thesection.\arabic{hran}}

\def\bmini{\setcounter{hran}{\value{equation}}
  \refstepcounter{hran}\setcounter{equation}{0}
  \renewcommand{\theequation}{\thehran\alph{equation}}\begin{eqnarray}}

\def\bminiG#1{\setcounter{hran}{\value{equation}}
\refstepcounter{hran}\setcounter{equation}{-1}
\renewcommand{\theequation}{\thehran\alph{equation}}
\refstepcounter{equation}\label{#1}\begin{eqnarray}}

\def\emini{\end{eqnarray}\relax\setcounter{equation
}{\value{hran}}\renewcommand{\theequation}{\thesection.\arabic{equation}}}

\def\beq{\begin{equation}}
\def\eeq{\end{equation}}
\newcommand{\be}{\begin{equation}}
\newcommand{\ee}{\end{equation}}
\def\bea{\begin{eqnarray}}
\def\eea{\end{eqnarray}}

\newcommand{\barre}[1]{%
	\setbox1=\hbox{$#1$} \dimen2=\ht1 \dimen3=\dp1 \dimen4=\wd1
	\setbox2=\hbox{\sl /}
	\dimen1=\wd1 \advance\dimen1 by -\wd2 \divide\dimen1 by 2
	\advance\dimen1 by \wd2 \advance\dimen1 by 0.4pt
	\setbox3=\hbox to \wd1{\hss \box1 \kern -\dimen1 \box2\hss}
	\ht3=\dimen2 \dp3=\dimen3 \wd3=\dimen4
	\box3
	}

\newcommand{\vev}[1]{%
	\langle #1 \rangle
	}

\def\1{{\rm 1 \kern -.10cm I \kern .14cm}} \def\R{{\rm R \kern -.28cm I
\kern .19cm}}

\begin{document}

\begin{titlepage}
\begin{flushright}    UFIFT-HEP-98-19 \\ 
\end{flushright}
\vskip 1cm
\centerline{\LARGE{\bf {The case for a}  }}
\vskip .5cm 
\centerline{\LARGE{\bf {Standard Model with Anomalous $U(1)$}}}
\vskip 1.5cm
\centerline{\bf Pierre Ramond}   
\vskip .5cm
\centerline{\em  Institute for Fundamental Theory,}
\centerline{\em Department of Physics, University of Florida}
\centerline{\em Gainesville FL 32611, USA}
\vskip 1.cm
\centerline{\it Invited Talk at Orbis Scientiae, Miami, FL, December
1977}
\vskip .5cm
\centerline{\bf {Abstract}}
\vskip .5cm
A gauged phase symmetry with its anomalies cancelled by a 
Green-Schwarz mechanism, broken at a large 
scale by an induced Fayet-Iliopoulos term, is a generic feature of a large class of 
superstring theories. It induces many desirable 
phenomenological features: Yukawa coupling hierarchy, the emergence of a 
small Cabibbo-like expansion parameter, relating the Weinberg angle to 
$b-\tau$ unification, and  the linking of 
R-parity conservation to neutrino masses. Some are discussed in the 
context of a three-family
model which reproduces all quark and lepton mass hierarchies as well
as the solar and atmospheric neutrino oscillations.
\vfill
\end{titlepage}


\mysection{Introduction}

The commonly accepted lore that string theories do not imply robust 
 relations among measurable parameters is challenged in a large class of 
effective low energy theories derived from string models contain an anomalous $U(1)$ with 
anomalies cancelled by the Green-Schwarz mechanism~\cite{GS} at cut-off. 
As emphasized by 't Hooft long ago, anomalies provide a link between 
infrared and ultraviolet physics. In these theories, this yields 
relations between the 
low-energy parameters of the standard model and those of the underlying 
theory through anomaly coefficients. An equally important feature is 
that as the dilaton 
gets a vacuum value, it generates a Fayet-Iliopoulos  that triggers the 
breaking~\cite{DSW} of the anomalous gauged symmetry at a  large 
computable scale.

Through the anomalous $U(1)$, the Weinberg angle at cut-off is 
related to anomaly coefficients~\cite{Ib}.  A simple model~\cite{IR} with one 
family-dependent anomalous $U(1)$ beyond the standard model was the first 
to exploit these features to produce Yukawa hierarchies and fix the 
Weinberg angle. It was soon 
realized that  some features could be abstracted from the 
presence of the anomalous $U(1)$: expressing the ratio of down-like quarks to 
charged lepton masses in terms of the Weinberg angle~\cite{BR,Nir,JS},
the suppression of the bottom to the top quark masses~\cite{PMR}, 
relating~\cite{BILR} the uniqueness of the vacuum 
to Yukawa hierarchies and the 
presence of MSSM invariants in the superpotential, and finally  
relating the see-saw mechanism~\cite{SEESAW} to R-parity 
conservation. 

These theories are expressed as effective low-energy supersymmetric 
theories with a cut-off scale $M$. The anomalous symmetry implies:

\begin{itemize}
\item A Cabibbo-like expansion parameter for the mass matrices.

\item Quark and charged lepton Yukawa hierarchies, and mixing, 
including the bottom to top Yukawa suppression. 

\item The value of the Weinberg angle at unification.   

\item Three massive neutrinos with mixings that give the small-angle 
MSW effect for the solar neutrino deficit, and the large angle mixing 
necessary for the atmospheric neutrino effect.

\item Natural R-parity conservation linked to massive 
neutrinos.

\item A hidden sector that contains  strong gauge interactions with 
chiral matter. 

\end{itemize}
An important theoretical requirement is that the vacuum, in which the 
anomalous symmetry is broken by stringy effects, be free of  flat directions 
associated with the MSSM invariants, and preserve supersymmetry.

The anomalous $U(1)$ also provides a possible explanation of supersymmetry 
breaking.  Since the  hidden sector contains a  
gauge theory with strong coupling,  the Green-Schwarz mechanism requires 
that it have a mixed anomaly as well. This implies that the hidden matter 
is chiral with respect to the anomalous symmetry. As shown by  
Bin\'etruy and Dudas~\cite{BD}, any strong coupling gauge theory with 
$X$-chiral fermions breaks supersymmetry (even QCD). A recent 
analysis~\cite{ADM} improves their mechanism by showing that the dilaton 
$F$-term does not vanish, providing for gaugino masses and possibly 
solving the  FCNC problem.

In the following, we present the generic features of this type of
model, and illustrate some in the context of a 
realistic three-family model.

\mysection{Applications to the standard model}

We consider models which have a gauge structure broken in two sectors: 
a visible sector, and a hidden sector, linked by the anomalous 
symmetry and possibly other Abelian symmetries (as well as gravity). 

\beq G_{\rm visible}\times U(1)_X\times U(1)_{Y^{(1)}}\cdots\times  
U(1)_{Y^{(M)}}\times G_{\rm hidden}\ ,\eeq  
where $G_{\rm hidden}$ is the hidden gauge group, and 

\beq G_{\rm visible}= SU(3)\times SU(2)\times U(1)_Y\ .\eeq  
is the standard model. Of the  $M+1$ extra symmetries,one  which we call $X$, 
is anomalous in the sense of 
Green-Schwarz.   

The  symmetries, $X$, $Y^{(a)}$ are spontaneously broken at a high 
scale by the Fayet-Iliopoulos term generated by the dilaton vacuum. This 
DSW vacuum~\cite{DSW} is required by phenomenology to preserve both supersymmetry and the 
standard model symmetries.

We assume the smallest matter content  needed to reproduce the 
observed quark and charged lepton hierarchy, cancel the anomalies 
associated with the extra gauge symmetries, and has a unique vacuum 
structure:
\begin{itemize}
\item Three chiral families
\item One standard-model vector-like pair of  Higgs 
weak doublets.
\item Three right-handed neutrinos $\overline N_i$, 
\item Standard model  vector-like pairs, 
\item Chiral fields that are needed to break 
the three extra $U(1)$ symmetries in the DSW vacuum. We denote these 
fields by $\theta_a$. 
\item Hidden sector gauge interactions and their matter, together with 
singlet fields, needed to cancel the remaining anomalies.
\end{itemize}


\mysection{Anomalies}

When viewed from the infrared, the anomaly constraints put 
strong restrictions on the low energy theory. In a four-dimensional theory, 
the Green-Schwarz anomaly compensation 
mechanism occurs through a dimension-five term that couples an axion to 
all the gauge fields. As a result, any anomaly linear in the 
$X$-symmetry must satisfy the Green-Schwarz relations

\be 
(XG_iG_j)=\delta_{ij}C_i\ ,\ee
where $G_i$ is any gauge current. The anomalous symmetry must have a 
mixed gravitational anomaly, so that 

\be
(XTT)=C_{\rm grav}\ne 0\ ,\ee
where $T$ is the energy-momentum tensor. In addition, the anomalies compensated 
by the Green-Schwarz mechanism satisfy the universality conditions
\be
{C_i\over k_i}\equiv{C_{\rm grav}\over 12}\ {\rm 
~~for~all~}~i\ .\ee
A similar relation holds for $C_X\equiv(XXX)$, the self-anomaly
coefficient of the $X$ 
symmetry.  These result in important numerical constraints, 
which restrict the matter content of the model. All other anomalies must vanish:
\be
(G_iG_jG_k)=(XXG_i)=0\ .\ee
In terms of the standard model, the vanishing anomalies are therefore of 
the following types: 
\begin{itemize}
 \item  The first involve only standard-model gauge groups $G_{\rm SM}$, 
with coefficients $(G_{\rm SM}G_{\rm SM}G_{\rm SM})$, which cancel for 
each chiral family and for vector-like matter. Also the hypercharge mixed 
gravitational anomaly $(YTT)$ vanishes.
\item The second type is where the new symmetries 
appear linearly, of the type $(Y^{(i)}G_{\rm SM}G_{\rm SM})$. If we 
assume that the 
$Y^{(i)}$ are traceless over the three chiral families, these vanish   over the 
three families of  fermions with standard-model charges. Hence they must vanish 
on the Higgs fields: with  $G_{\rm SM}=SU(2)$, it implies the Higgs pair is 
vector-like  with respect to the $Y^{(i)}$. It also follows that the mixed 
gravitational anomalies $(Y^{(i)}TT)$ are zero over the fields with 
standard model quantum numbers. 
 
\item The third type involve the new symmetries quadratically, of the form $(G_{\rm 
SM}Y^{(i)}Y^{(j)})$. These vanish by group theory except for those 
 of the form $(YY^{(i)}Y^{(j)})$. In general two types of 
fermions contribute: the three chiral families and 
standard-model vector-like pairs. 

\item The remaining vanishing anomalies involve the anomalous charge $X$. 
\begin{itemize} 
\item With $X$ family-independent, and 
$Y^{(i)}$ family-traceless,  the vanishing of the 
$(XYY^{(i)})$ anomaly coefficients   over the three families is
assured:  so 
they must also vanish over the Higgs 
pair. This means that $X$ is vector-like on the Higgs pair. 
This is an important result, as it implies that the standard-model invariant $H_uH_d$ (the $\mu$ 
term) has zero $X$ and $Y^{(i)}$ charges; it can appear by itself 
in the superpotential, but we are dealing with a 
string theory, where mass terms do not appear in the superpotential: it 
can appear only in the K\"ahler potential. This results, after 
supersymmetry-breaking in 
an induced $\mu$-term, of weak strength, as suggested by Giudice and 
Masiero~\cite{GM}. 

\item The  coefficients $(XY^{(i)}Y^{(j)})$, $i\ne j$. Since 
standard-model singlets can contribute to these anomalies, we expect 
cancellation to 
come about through a combination of hidden sector and singlet fields.
\item The coefficient $(XXY)$. This imposes an important constraint on 
the $X$ charges on the chiral families.  
\item The coefficients $(XXY^{(i)})$; with family-traceless symmetries, 
they  vanish over the three families of 
fermions with standard-model charges, but contributions are expected from 
other sectors of the theory.
\end{itemize}
\end{itemize}
The building of models in which these anomaly coefficients vanish is 
highly non-trivial. Finding a set of charges which satisfy all anomaly 
constraints, and reproduce phenomenology is highly constrained. In the 
three-family model it will even prove predictive in the neutrino sector. 

\subsection{Standard Model Anomalies}
In the standard model, we consider three anomalies associated with its 
three gauge groups, 

\be 
C_{\rm color}=(XSU(3)SU(3))\ ;\qquad C_{\rm weak}=(XSU(2)SU(2))\ ;\qquad 
C_{\rm Y}=(XYY)\ ,\ee
when $()$ stands for the trace. They can be expressed~\cite{PMR}
 in terms of the $X$-charges of 
the invariants of the MSSM

\be
C^{}_{\rm color}={1\over 2}\sum_{i}\left[X^{[u]}_{ii}+X^{[d]}_{ii}\right]
-3X^{[\mu]}_{}\ ,\ee

\be
C_{\rm Y}+C_{\rm weak}-{8\over 3}C_{\rm 
color}=2\sum_i\left[X^{[e]}_{ii}-X^{[d]}_{ii}\right]
+2X^{[\mu]}_{}\ ,\ee
where $X^{[u]}_{ij}$ is the $X$-charge of ${\bf Q}^{}_i{\bf\overline 
u}^{}_jH^{}_u$, $X^{[d]}_{ij}$ that of ${\bf Q}^{}_i{\bf\overline 
d}^{}_jH^{}_d$, $X^{[e]}_{ij}$ that of $L^{}_i{\overline e}^{}_jH^{}_d$, 
and finally
$X^{[\mu]}$ that of the $\mu$-term $H^{}_uH^{}_d$, 
where $i,j$ are the family indices. 
Also the mixed gravitational anomaly over the three chiral families is 
given by

\be
C_{\rm grav}^{[\rm fam]}=3C_{\rm color}+\sum_iX^{[e]}_{ii}
-X_{}^{[\mu]}\ .\ee

In theories derived from superstrings, the integer level numbers $k_{\rm color}$ 
and $k_{\rm weak}$ are equal, resulting in the equality

\be
C_{\rm weak}=C_{\rm color}\ .\ee

These imply that, as long as these anomaly coefficients do 
not vanish, the MSSM Yukawa invariants cannot all appear at tree level, as their 
$X$-charges are necessarily non-zero. This means that not all 
Yukawa couplings can be of the same order of magnitude, resulting in some 
sort of Yukawa hierarchy.
 
More specific conclusions can be reached by assuming that the $X$ charges 
are family-independent and the $Y^{(i)}$ are family-traceless. As we have 
seen, the 
$\mu$-term has vector-like charges, $X^{[\mu]}=0$. 

By further assuming that the top quark Yukawa mass coupling occurs at 
tree-level, we have $X^{[u]}_{33}=X^{[u]}_{}=0$. This implies that the 
$X$-charge of the down quark Yukawa is proportional to the color anomaly, 
and  thus cannot vanish: the down Yukawa is {\it necessarily} smaller 
than the top Yukawa, leading to the suppression of $m_b$ over $m_t$, 
after electroweak breaking! The non-vanishing of the color anomaly 
implies the (observed) suppression of the bottom mass relative to the top 
mass.

The second anomaly equation simplifies to  

\be
C_{\rm Y}-{5\over 3}C_{\rm weak}=6\left[X^{[e]}_{}-X^{[d]}_{}\right]
\ ,\ee
stating that the relative suppression of the 
down to the charged lepton sector is proportional to the difference of 
two anomaly coefficients. The data, extrapolated to near unification 
scales indicates that there is no relative suppression between the two 
sectors, suggesting that difference should vanish. Remarkably, the 
vanishing~\cite{Ib} of that combination fixes the value of the Weinberg angle
through the string of relations

\be
{3\over 5}={C_{\rm weak}\over C_{\rm Y}}={k_{\rm weak}\over k_{\rm Y}}={g^2_Y\over 
g^2_{\rm weak}}=\tan^2\theta_w\ .\ee
This happens exactly at the phenomenologically 
preferred value of the Weinberg angle: the $b-\tau$ unification is
related to the value of the Weinberg angle~\cite{BR}!

The application of the Green-Schwarz structure to 
the standard model is consistent with many of its phenomenological 
patterns. However, more can be said through a careful study of the DSW 
vacuum.


\mysection{The DSW vacuum}
\label{section:FD}

When the dilaton acquires its vacuum value, an anomalous Fayet-iliopoulos 
$D$-term is generated through the gravitational anomaly. In the weak 
coupling limit of the string, it is given by 

\be
\xi^2={g^2\over 192\pi^2}M^2_{\rm Planck}C_{\rm grav}\ ,\ee
where $g$ is the string coupling constant. This induces the breaking of
 $X$ and $Y^{(i)}$ 
below the cut-off. 
 
Phenomenology require that neither 
supersymmetry nor any of the standard model symmetries be broken at that 
scale. This puts severe restrictions on the form of the superpotential 
and the  matter fields \cite{BILR}. 

The analysis of the vacuum structure of supersymmetric theories is 
greatly facilitated by the fact that the solutions of the vacuum 
equations for the D-terms are in one-to-one correspondance with 
holomorphic invariants. This analysis has been recently generalized to 
include an anomalous Fayet-Iliopoulos term. 

In order to get a unique determination of the DSW vacuum, we need 
as many singlet superfields, $\theta_a$,  as there are broken 
symmetries. Only they  assume vev's as a result 
of the FI term. They are 
standard model singlets, but not under $X$ and 
$Y^{(a)}$. If more 
fields than broken symmetries assume non-zero values in the DSW vacuum, 
we would have undetermined flat directions and hierarchies.

We assemble the charges in a $(M+1)\times (M+1)$ matrix $\bf A$, whose 
rows are the $X$, $Y^{(i)}$ charges of the $\theta$ fields, respectively. 
Assuming the existence of a supersymmetric  vacuum where only the $\theta$
fields have vacuum values, implies from the vanishing of the $M+1$ $D$-terms 

\be 
{\bf 
A}\left(\begin{array}{c}\vert\theta_1\vert^2\\\cdot\\
\vert\theta_{M+1}\vert^2
\end{array}  \right)=\left(\begin{array}{c}\xi^2\\\cdot\\ 
0\end{array}\right)\ .\ee
For this vacuum solution to exist, the matrix $\bf A$ must have 
an inverse and the  entries in the first row of its inverse must be positive. 
The solution to these equations naturally 
provide computably small expansion parameters
$\lambda_a=<\vert\theta_a\vert>_0/M$. In the case when all expansion
parameters are the same we can relate their value in terms of standard
model quantities

\be
\lambda_a={\alpha\over 4\pi}C_{\rm color}\ ,\ee
where $\alpha$ is the unified gauge coupling at unification. 

Another  important consequence is that there is no holomorphic invariant 
polynomial involving the $\theta$ fields alone. Another is that the 
$\theta$ sector is necessarily anomalous. Indeed, let us assume that it 
has no mixed gravitational anomalies. This means that all the charges are 
traceless over the $\theta$ fields. now the $(M+1)$ $\theta$ fields form a 
representation of $SU(M+1)$, and the tracelessness of the charges insures 
that they be members of $SU(M+1)$. So we are looking for $M$ 
non-anomalous symmetries in $SU(M+1)$, which is impossible except for 
$M=1$. If two or more of the charges are the same on the $\theta$'s, we 
could have anomaly cancellation, but then the matrix $A$ would be 
singular, contrary to the assumption of the DSW vacuum. 
Hence this sector will in general be anomalous.

For a thorough analysis of the vacuum with FY term, we refer the reader 
to Ref. ~\cite{IL,BILR}. Here, we simply note two striking generic facts of 
phenomenological import. Consider any  invariant $I$ of the MSSM. It 
corresponds~\cite{Buccella} to a possible flat direction of the non-anomalous 
supersymmetric vacuum. For that configuration, all its fields are aligned  
to the same vacuum value, as required by the vanishing of 
the non-anomalous $D$-terms of the standard model symmetries.
It follows that the contribution of these terms of the anomalous $D_X$ 
will be proportional to its $X$-charge~\cite{DGPS}. In order to forbid 
this flat direction to appear alone in the vacuum, it is therefore necessary to 
require that its charge be of the  wrong sign to forbid a solution of 
$D_X=\xi^2$. This implies a holomorphic invariant of the form $I{\cal 
P}(\theta_a)$, where $\cal P$ is a holomorphic polynomial in the 
$\theta$'s. The $D$-term equations are not sufficient to forbid this flat 
direction together with $\theta$ fields. We have to rely on the $F$-terms 
associated with that invariant polynomial, and its presence is needed in 
the superpotential. Fortunately,  phenomenology {\bf also} 
requires such terms to appear in 
the superpotential. This is the first of several curious links between 
phenomenology and the vacuum structure near unification scales! One can 
see that the existence of this invariant is predicated on the  invertibility of
 ${\bf A}$, the same condition for the DSW vacuum.

The second point addresses singlet fields that do not get vev's in the 
DSW vacuum. To implement the seesaw mechanism, there must be right-handed 
neutrinos, $\overline N_i$ .  Since they have no vev, their X-charge must 
also be of the wrong sign, which allows for holomorphic invariants of the 
form $\overline 
N^A{\cal P}(\theta)$, where $A$ is a positive integer. The case $A=1$ is 
forbidden as it breaks supersymmetry. Thus $A\ge 2$. The case $A=2$ 
generates Majorana masses for these fields in the DSW vacuum. W 
scale. To single out  
$A=2$ we need to choose the $X$ charges 
of the $\overline N_i$ to be a negative half-odd integers.  To implement 
the seesaw, the right-handed neutrinos couple to the standard model invariants
$L_iH_u$, which requires that $X_{L_iH_u}$ is also a half-odd
integer, while all other MSSM invariants have positive or zero integers 
$X$-charges.


\mysection{A Three-Family Model}
We can see how some of the features we have just discussed lead to
phenomenological consequences in the context of 
a three-family model~\cite{EIR,ILR}, 
with three Abelian symmetries broken in the DSW vacuum. The matter
content of the theory is inspired by $E_6$, which 
 contains two Abelian symmetries outside of the 
standard model: the first $U(1)$, which we call $V'$, appears in the embedding 
\be E_6\subset ~SO(10)\times U(1)_{V'}\ee
 The second $U(1)$, called $V$, appears in
\be SO(10)\subset SU(5)\times U(1)_V\ .\ee
The two non-anomalous symmetries are  

\beq Y^{(1)}={1\over 5}(2Y+V)
\left( \begin{array}{ccc}
    2 & 0 & 0 \\
    0 & -1 & 0 \\
    0 & 0 & -1   \end{array}  \right)\eeq

\beq Y^{(2)}={1\over 4}(V+3V')\left( \begin{array}{ccc}
    1 & 0 & 0 \\
    0 & 0 & 0 \\
    0 & 0 & -1   \end{array}\right)\ ,\eeq
The family matrices run over the three chiral families, so that $Y^{(1,2)}$ 
are family-traceless. Since ${\rm Tr}(YY^{(i)})=0$, there is no appreciable
    kinetic mixing between the non-anomalous $U(1)$s.

The $X$ charges on the three chiral families in the $\bf 
27$ are of the form

\be X=(\alpha+\beta V+\gamma V')\left( \begin{array}{ccc}
    1 & 0 & 0 \\
    0 & 1 & 0 \\
    0 & 0 & 1   \end{array}\right)\ ,\eeq
where $\alpha,~\beta,~\gamma$ are expressed in terms of the
    $X$-charges of $\overline N_i$ (=-3/2), that of ${\bf Q}\overline{\bf
    d }H_d$ (=-3), and that of the vector-like pair ,mass term
    $\overline E E$ (=-3).
 
\noindent The matter content of this model is the smallest that reproduces the 
observed quark and lepton hierarchy while cancelling the anomalies 
associated with the extra gauge symmetries:
\begin{itemize}
\item Three chiral families each with the quantum numbers of 
a $\bf 27$ of $E_6$. This means  three chiral families of the standard 
model, ${\bf Q}_i$, $\overline{\bf u}_i$, $\overline{\bf d}_i$, $L_i$, 
and $\overline e_i$, together with three right-handed neutrinos $\overline N_i$, 
three vector-like pairs denoted by $E_i$ + $\overline{\bf D}_i$ 
and $\overline E_i$ + ${\bf D}_i$, with the quantum numbers of the $\bf 5$ + 
$\overline{\bf 5}$ of $SU(5)$, and finally three real singlets $S_i$. 
\item One standard-model vector-like pair of  Higgs 
weak doublets.
\item Chiral fields that are needed to break 
the three extra $U(1)$ symmetries in the DSW vacuum. We denote these 
fields by $\theta_a$. In our minimal model with three symmetries that 
break through the FI term, we just take $a=1,2,3$. The 
$\theta$ sector  is necessarily anomalous.
\item Other standard model singlet fields.
\item Hidden sector gauge interactions and their matter.
\end{itemize}

Finally, the charges of the $\theta$ fields is given in terms of the matrix
\be
{\bf A}= \left( \begin{array}{ccc}
1&0&0\\ 0&-1&1\\ 1&-1&0
\end{array}  \right)\ .
\ee
Its inverse 
\be
{\bf A}^{-1}= \left( \begin{array}{ccc}
1&0&0\\ 1&0&-1\\ 1&1&-1
\end{array}  \right)\ ,\ee
shows all three fields acquire the same vacuum value. 

In the following, we will address only the features of the model which
are of more direct phenomenological interest. For more details, the
interested reader is referred to the original references ~\cite{EIR,ILR}.

\subsection{Quark and Charged Lepton Masses}

The Yukawa interactions in the charge $2/3$ quark sector are generated
by operators of the form
\be {\bf Q}^{}_i\bar{\bf u}^{}_jH^{}_u
{\bigl ( {\theta_1 \over M} \bigr )}^{n^{(1)}_{ij}}
{\bigl ( {\theta_2 \over M} \bigr )}^{n^{(2)}_{ij}}
{\bigl ( {\theta_3 \over M} \bigr )}^{n^{(3)}_{ij}}
\ ,\label{eq:uterm}\ee
in which the exponents must be positive integers or zero. Assuming
that only the top quark Yukawa coupling appears at tree-level, a
straighforward computation of their charges yields in the DSW vacuum
the charge $2/3$ Yukawa matrix 
 \be
Y_{}^{[u]}\sim\left( \begin{array}{ccc}
\lambda^8 &\lambda^5&\lambda^3\\ \lambda^7&\lambda^4&\lambda^2\\
\lambda^5&\lambda^2&1\end{array}  \right)\ ,\ee
where $\lambda=\vert\theta_a\vert/ M$ is the common expansion parameter.

 A similar computation is now applied to the charge $-1/3$ Yukawa standard 
model invariants ${\bf Q}^{}_i\bar{\bf d}^{}_jH^{}_d$. The difference is 
that  $X^{[d]}$, its $X$-charge does not vanish. As long as
$X^{[d]}\le -3$, we deduce the charge $-1/3$ Yukawa matrix  

\be 
Y_{}^{[d]}\sim\lambda_{}^{-3X^{[d]}-6}\left( \begin{array}{ccc}
\lambda_{}^{4} &\lambda_{}^{3}&\lambda_{}^{3}\\ 
\lambda_{}^{3}&\lambda_{}^{2}&\lambda_{}^{2}\\
\lambda_{}^{}&1&1\end{array}  \right)\ .\ee
Diagonalization of the two Yukawa matrices yields the 
CKM matrix
\be
{\cal U}^{}_{CKM}\sim\left( \begin{array}{ccc}
1 &\lambda_{}^{}&\lambda_{}^{3}\\ \lambda_{}^{}&1&\lambda_{}^{2}\\
\lambda_{}^{3}&\lambda_{}^{2}&1\end{array}  \right)\ .\ee
This shows  the expansion parameter to be of the same order of magnitude 
as the Cabibbo angle $\lambda_c$.

The eigenvalues of these matrices reproduce the 
geometric interfamily hierarchy for quarks of both charges

\be
{m_u\over m_t}\sim \lambda_c^8\ ,\qquad {m_c\over m_t}\sim
\lambda_c^4\ .\ee
\be
{m_d\over m_b}\sim\lambda_c^4\ ,\qquad {m_s\over m_b}\sim \lambda_c^2\
,\ee
while the quark intrafamily hierarchy is given by
\be
{m_b\over m_t}= \cot\beta\lambda_{c}^{-3X^{[d]}-6}\ .\ee
implying the relative suppression of the bottom to  top quark masses, 
without large $\tan\beta$. 
These quark-sector results are the same as in a previously published 
model~\cite{EIR}, but our present model is different in the lepton 
sector. 

The analysis in the charged lepton sector proceeds in similar ways. 
No dimension-three term appears and the standard model invariant 
$L_i\overline e_jH_d$ have $X$-charge $X^{[e]}$.  For  $X^{[e]}=-3$, there are 
supersymmetric zeros in the $(21)$ and $(31)$ position, yielding

\be 
Y_{}^{[e]}\sim\lambda_{c}^{3}\left( \begin{array}{ccc}
\lambda_c^{4} &\lambda_c^{5}&\lambda_c^{3}\\ 
0&\lambda_c^{2}&1\\
0&\lambda_c^2&1\end{array}  \right)\ .\ee
Its diagonalization yields the lepton interfamily hierarchy

\be
{m_e\over m_\tau}\sim\lambda_c^4\ ,\qquad {m_\mu\over m_\tau}\sim
\lambda_c^2\ .\ee
Our choice of $X$ insures that $X^{[d]}=X^{[e]}$, which guarantees 
through the anomaly conditions the 
correct value of the Weinberg angle at cut-off, since 

\be \sin^2\theta_w={3\over 8}~~~\leftrightarrow ~~~X^{[d]}=X^{[e]}\ ;\ee
it  sets $X^{[d]}=-3$, so that  

\be
{m_b\over m_\tau}\sim 1\ ;\qquad {m_b\over m_t}\sim 
\cot\beta\lambda_c^3\ .\ee
It is a remarkable feature of this type of model that both inter- and 
intra-family hierarchies are linked not only with one another but with 
the value of the Weinberg angle as well. In addition, the model predicts
a natural suppression of $m_b/m_\tau$, which suggests that $\tan \beta$
is of order one.


\subsection{Neutrino Masses}
Neutrino masses are naturally generated by the seesaw mechanism~\cite{SEESAW}
if the three right-handed neutrinos $\overline N_i$ acquire a Majorana mass
in the DSW vacuum. The flat direction analysis indicates that their
$X$-charges must be negative half-odd integers, with 
 $X_{\overline N}=-3/2$ preferred by the vacuum analysis. 
Their standard-model invariant masses are generated by terms of the form

\be 
M\overline N_i\overline N_j
{\bigl ( {\theta_1 \over M} \bigr )}^{p^{(1)}_{ij}}
{\bigl ( {\theta_2 \over M} \bigr )}^{p^{(2)}_{ij}}
{\bigl ( {\theta_3 \over M} \bigr )}^{p^{(3)}_{ij}}\ ,\ee
where $M$ is the cut-off of the theory. In the $(ij)$ matrix element. 
The Majorana mass matrix is computed to be 

\be
M\lambda_c^{7}\left( \begin{array}{ccc}
\lambda_c^{6} &\lambda_c^{5}&\lambda_c^{}\\ 
\lambda_c^{5}&\lambda_c^{4}&1\\
\lambda_c^{}&1&0\end{array}  \right)\ .\ee
Its diagonalization yields three massive right-handed neutrinos with masses 

\be
m_{\overline N_e}\sim M\lambda_c^{13}\ ;\qquad m_{\overline N_\mu}\sim 
m_{\overline N_\tau}\sim M\lambda_c^7\ .\ee

By definition, right-handed neutrinos are those that 
couple to the standard-model invariant $L_iH_u$, 
and serve as Dirac partners to the chiral neutrinos. In our model,

\be
X(L_iH_u\overline N_j)\equiv X^{[\nu]}=0\ .\ee
The superpotential contains the terms 

\be
L_iH_u\overline N_j
{\bigl ( {\theta_1 \over M} \bigr )}^{q^{(1)}_{ij}}
{\bigl ( {\theta_2 \over M} \bigr )}^{q^{(2)}_{ij}}
{\bigl ( {\theta_3 \over M} \bigr )}^{q^{(3)}_{ij}}\ \ee
resulting, after electroweak symmetry breaking, in the orders of magnitude
(we note $v_u = \vev{H^0_u}$)

\be
 v_u  \left( \begin{array}{ccc}
\lambda_c^{8}&\lambda_c^{7}&\lambda_c^{3}\\  \lambda_c^{5}&\lambda_c^{4}&1
\\ \lambda_c^{5}&\lambda_c^{4}&1\end{array}  \right)\ \ee
for the neutrino Dirac mass matrix. The actual 
neutrino mass matrix is generated by the seesaw mechanism. 
A careful calculation yields the orders of magnitude

\be
{v_u^2\over M\lambda_c^3} \left( \begin{array}{ccc}
\lambda_c^6&\lambda_c^3&\lambda_c^3\\ \lambda_c^3&1&1\\ \lambda_c^3&1&1
\end{array}  \right)\ .
\label{eq:nu_matrix}\ee
A characteristic of the seesaw mechanism is that the 
charges of the $\overline N_i$ do not enter in the determination of these 
orders of magnitude as long as there are no massless right-handed
neutrinos. Hence the structure of the neutrino mass matrix depends only on
the charges of the invariants $L_iH_u$, already fixed by phenomenology and
anomaly cancellation. In particular, the family structure is that
carried by the lepton doublets $L_i$. In our model, since 
$L_2$ and $L_3$ have the same charges, it ias not surprising that we
 have no flavor
distinction between the neutrinos of the second and third family. 
In models with two non-anomalous flavor 
symmetries based on $E_6$ the matrix (\ref{eq:nu_matrix}) is a very stable
prediction of our model. Its diagonalization yields the neutrino mixing
matrix~\cite{MNS}

\be
 {\cal U}_{\rm MNS}=\left( \begin{array}{ccc}
1&\lambda_c^{3}&\lambda_c^{3}\\  \lambda_c^{3}&1&1
\\ \lambda_c^{3}&1&1\end{array}  \right)\ ,\ee
so that the mixing of the electron neutrino is small, of the order of
$\lambda_c^3$, while the mixing between the $\mu$ and $\tau$ neutrinos is of
order one. Remarkably enough, this mixing pattern is precisely the one
suggested by the non-adiabatic MSW ~\cite{MSW} explanation of the solar
neutrino deficit and by the oscillation interpretation of the reported
anomaly in atmospheric neutrino fluxes (which has been recently confirmed
by the Super-Kamiokande \cite{SuperK} and Soudan \cite{Soudan}
collaborations). 

Whether the present model actually fits both 
the experimental data on solar and
atmospheric neutrinos or not depends on the eigenvalues of the mass matrix
(\ref{eq:nu_matrix}). A naive order of magnitude diagonalization gives a
$\mu$ and $\tau$ neutrinos of comparable masses, and a much lighter electron
neutrino:
\be
m_{\nu_e}\ \sim\ m_0\, \lambda_c^{6}\ ;\qquad 
m_{\nu_\mu},\, m_{\nu_\tau}\ \sim\ m_0\ ;\qquad
m_0\ =\ {v_u^2\over M\lambda_c^3}\ ,
\label{eq:nu_mass}\ee
The overall neutrino mass scale $m_0$ depends on the cut-off $M$. Thus the
neutrino sector allows us, in principle, to measure it.

At first sight, this spectrum is not compatible with a simultaneous
explanation of the solar and atmospheric neutrino problems, which requires
a hierarchy between $m_{\nu_\mu}$ and $m_{\nu_\tau}$. However, the estimates
(\ref{eq:nu_mass}) are too crude: since the (2,2), (2,3) and (3,3) entries
of the mass matrix all have the same order of magnitude, the prefactors that
multiply the powers of $\lambda_c$ in (\ref{eq:nu_matrix}) can spoil the
naive determination of the mass eigenvalues. A more careful analysis 
shows that even with factors of order one, it is possible to fit the 
atmospheric neutrino anomaly as well. A welcome by-product of the 
analysis is that the mixing angle is actually driven to its maximum 
value. We refer the reader to Ref.\cite{ILR} for more details. The main 
point of this analysis is that maximal mixing between the second and 
third family in the neutrino sector occurs naturally as it is determined 
from the structure of the quark and charged lepton hierarchies.



\mysection{R-Parity}

The invariants of the minimal standard model and their associated flat 
directions have been analyzed in detail in the literature~\cite{GKM}. In 
models with an anomalous $U(1)$, these invariants carry in general  
$X$-charges, which,  as we have seen,  determines their suppression in 
the effective Lagrangian. Just as there is a basis of invariants,  proven 
long ago by Hilbert, the charges of these invariants are not all 
independent; they can in fact be expressed in terms of the charges of the 
lowest order invariants built out of the fields of the minimal standard 
model, and some anomaly coefficients. 

The $X$-charges of the three types of cubic standard model invariants
that violate 
$R$-parity as well as baryon and/or lepton numbers can be expressed in 
terms of the $X$-charges of the MSSM invariants and the R-parity 
violating invariant
 
\be
  X^{[\barre R]}\equiv X(LH_u)\ ,\ee
through the relations

\be
X_{L{\bf Q}\bar{\bf d}}=X^{[d]}-X^{[\mu]}+X^{[\barre R]}\ 
,\nonumber\ee
 \be
X_{LL\bar e}=X^{[e]}-X^{[\mu]}+X^{[\barre R]}\ .\nonumber\ee
\be X_{\bar{\bf u}\bar{\bf d}\bar{\bf d}}=X^{[d]}+X^{[\barre R]}
+{1\over 3}\left(C_{\rm color}-C_{\rm weak}\right)-{2\over 3}X^{[\mu]}\ .\nonumber\ee
Although they vanish in our model, we still display 
 $X^{[u]}$ and $X^{[\mu]}=0$, since these sum rules are more general. 

In the analysis of the flat directions, we have seen how 
the seesaw mechanism forces the $X$-charge of 
$\overline N$ to be half-odd integer. Also, the Froggatt-Nielsen~\cite{FN} 
suppression of the minimal standard model invariants, and the holomorphy 
of the superpotential  require 
$X^{[u,d,e]}$ to be zero or negative integers, and the equality of the 
K\'ac-Moody levels of $SU(2)$ and $SU(3)$  forces $C_{\rm color}=C_{\rm 
weak}$, through the Green-Schwarz mechanism. Thus we conclude that the 
$X$-charges of these operators are half-odd integers, and thus they cannot 
appear in the superpotential unless multiplied by at least one 
$\overline N$. This reasoning can be applied to the  higher-order 
$\barre R$ operators since their charges are given by 

\bea
X_{{\bf Q}{\bf Q}{\bf Q}H_d}&=&
X^{[u]}+X^{[d]}-{1\over 3}X^{[\mu]}-X^{[\barre R]}\ ,\\ 
X_{\bar{\bf d}\bar{\bf d}\bar{\bf d}LL}&=&
2X^{[d]}-X^{[u]}-{5\over 3}X^{[\mu]}+3X^{[\barre R]}\ ,\\
X_{{\bf Q}{\bf Q}{\bf Q}{\bf Q}\bar{\bf u}}&=&
2X^{[u]}+X^{[d]}-{4\over 3}X^{[\mu]}-X^{[\barre R]}\ ,\\ 
X_{\bar{\bf u}\bar{\bf u}\bar{\bf u}\bar e\bar e}&=&
2X^{[u]}-X^{[d]}+2X^{[e]}-{2\over 3}X^{[\mu]}-X^{[\barre R]}\ ,
\eea
It follows that {\bf there are no $R$-parity violating operators, 
whatever their dimensions} : through the right-handed 
neutrinos, $R$-parity is linked to half-odd integer charges, so that 
charge invariance results in $R$-parity invariance. Thus  {\bf none} of 
the operators that violate $R$-parity can appear in holomorphic invariants: 
even after breaking of the anomalous $X$ symmetry, the remaining 
interactions all respect $R$-parity, leading to an {\bf absolutely 
stable superpartner}.  This is a general result deduced from  
the uniqueness of the DSW vacuum, the 
Green-Schwarz anomaly cancellations, and the seesaw mechanisms.


\mysection{Conclusion}
The case for an anomalous $U(1)$ extension to the standard model is
particularly strong. We have presented many of its phenomenological
consequences. In a very unique model, we detailed how the neutrino
matrices are predicted. However much remains to be done: the nature of
the hidden sector, and supersymmetry breaking. Our model only predicts
orders of magnitude of Yukawa couplings. To calculate the prefactors,
a specific theory is required. Many of the features we have discussed
are found in the context of free fermion theories~\cite{faraggi}, which
arise in the context of perturbative string theory. It is hoped that
since they involve anomalies, these features can also be derived under
more general assumptions. A particularly difficult problem is that of
the cut-off scale. From the point of view of the low energy, there is
only one scale of interest, that at which the couplings unify, and the
Green-Schwarz mechanism, by fixing the weak and color anomalies,
identifies the cut-off as the unification scale. On the other hand,
another mass scale appears in the theory through the size of the
anomalous FI term, and the two values do not coincide, the usual
problem of string unification. It is hoped that the calculation of the
Fayet-Iliopoulos term in other regimes will throw some light on this
problem. 
 
Our simple model has too many desirable phenomenological
features to be set aside, and we hope that a better understanding of
fundamental theories will shed light on this problem. 
\vskip 1cm 
{\bf Acknowledgements}
\vskip .5cm

I would like to thank Professor B. Kursunoglu for his kind hospitality 
and for giving me the opportunity to speak at this pleasant
conference, as well as my collaborators, N. Irges and S. Lavignac, 
on whose work much of the above is based. This work was supported in part by the
United States Department of Energy under grant DE-FG02-97ER41029.


\end{document}